\documentclass[aps,prl,twocolumn,numerical,superscriptaddress,nofootinbib,showpacs,showkeys,longbibliography]{revtex4-1}

\usepackage{graphicx,color}
\usepackage{amsmath,amssymb,amsfonts}

\newcommand{\be}{\begin{equation}}
\newcommand{\ee}{\end{equation}}
\newcommand{\ba}{\begin{eqnarray}}
\newcommand{\ea}{\end{eqnarray}}
\newcommand{\ban}{\begin{eqnarray*}}
\newcommand{\ean}{\end{eqnarray*}}

\def\v2{\mbox{$v_2$}}

\def\sqrtsNN{\mbox{$\sqrt{s_{\mathrm{NN}}}$}}

\begin{document}

\title{An extended $R^{(2)}_{\Psi_{m}}(\Delta S_{2})$ correlator for detecting and characterizing \\ the Chiral Magnetic Wave}
\medskip

\author{Niseem~Magdy} 
\email{niseemm@gmail.com}
\affiliation{Department of Physics, University of Illinois at Chicago, Chicago, Illinois 60607, USA}

\author{Mao-Wu Nie}
\affiliation{Institute of Frontier and Interdisciplinary Science, Shandong University, Qingdao, Shandong, 266237, China}
\affiliation{Key Laboratory of Particle Physics and Particle Irradiation, Ministry of Education, Shandong University, 
Qingdao, Shandong, 266237, China}

\author{Ling Huang}
\affiliation{Shanghai Institute of Applied Physics, Chinese Academy of Sciences, Shanghai 201800, China}
\affiliation{University of Chinese Academy of Sciences, Beijing 100049, China}
\affiliation{Key Laboratory of Nuclear Physics and Ion-beam Application (MOE), Institute of Modern Physics, Fudan University, Shanghai 200433, China}

\author{Guo-Liang Ma}
\email[]{glma@fudan.edu.cn}
\affiliation{Shanghai Institute of Applied Physics, Chinese Academy of Sciences, Shanghai 201800, China}
\affiliation{Key Laboratory of Nuclear Physics and Ion-beam Application (MOE), Institute of Modern Physics, Fudan University, Shanghai 200433, China}

\author{Roy~A.~Lacey} 
\email{Roy.Lacey@stonybrook.edu}
\affiliation{Depts. of Chemistry \& Physics, Stony Brook University, Stony Brook, New York 11794, USA}

\date{\today}

\begin{abstract}
The extended $R^{(2)}_{\Psi_{m}}(\Delta S_{2})$ correlator is presented and examined for its efficacy 
to detect and characterize the quadrupole charge separation ($\Delta S_{2}$) associated with the purported 
Chiral Magnetic Wave (CMW) produced in heavy-ion collisions. 
Sensitivity tests involving varying degrees of proxy CMW signals injected into events simulated 
with the Multi-Phase Transport Model (AMPT), show that the $R^{(2)}_{\Psi_{m}}(\Delta S_{2})$ 
correlator provides discernible responses for background- and CMW-driven charge separation. 
This distinction could aid  identification of the CMW via measurements of the $R^{(2)}_{\Psi_{2}}(\Delta S_{2})$ and 
$R^{(2)}_{\Psi_{3}}(\Delta S_{2})$ correlators, relative to the second- ($\Psi_{2}$) and third-order ($\Psi_{3}$) event planes.  
The tests also indicate a level of sensitivity that would allow for robust experimental 
characterization of the CMW signal.

\end{abstract}

\maketitle

Heavy-ion collisions at the Relativistic Heavy Ion Collider (RHIC) and the Large Hadron Collider (LHC) can 
lead to a magnetized chiral relativistic quark-gluon plasma (QGP) \cite{Kharzeev:2004ey,Liao:2014ava,Miransky:2015ava,
Huang:2015oca,Kharzeev:2015znc}, in which the mass of fermions are negligible compared to the 
temperature and/or chemical potential. Such a plasma, which  is akin to the primordial plasma in the 
early Universe \cite{Rogachevskii:2017uyc,Gorbunov:2011zz} and several types of degenerate forms 
of matter in compact stars \cite{Weber:2004kj}, have pseudo-relativistic analogs in 
Dirac and Weyl materials \cite{Vafek:2013mpa,Burkov:2015hba,Gorbar:2017lnp}. 
It is further characterized not only by an exactly conserved electric charge but also by an approximately 
conserved chiral charge, violated only by the quantum chiral anomaly \cite{Adler:1969gk,Bell:1969ts}.

 The study of anomalous transport in magnetized chiral plasmas can give fundamental insight not only 
on the complex interplay of chiral symmetry restoration, axial anomaly and gluon topology 
in the QGP \cite{Moore:2010jd,Mace:2016svc,Liao:2010nv,Kharzeev:2015znc,Skokov:2016yrj}, but also on 
the evolution of magnetic fields in the early Universe \cite{Joyce:1997uy,Tashiro:2012mf}. 
Two of the principal anomalous processes in these plasmas 
[for electric and chiral charge chemical potential $\mu_{V,A} \neq 0$] 
are the chiral separation effect (CSE) \cite{Vilenkin:1980fu,Metlitski:2005pr,Son:2009tf} 
and the chiral magnetic effect (CME) \cite{Fukushima:2008xe}.
The CSE is derived from the induction of a non-dissipative chiral axial current:
\begin{equation}
\vec{J}_A = \frac{e\vec{B}}{2\pi^2}\mu_V, {\rm for}\, \mu_V\neq 0,
\end{equation}
where $\mu_V$ is the vector (electric) chemical potential and $\vec{B}$ is the magnetic field. 
The CME is similarly characterized  by the vector current:
\begin{equation}
\vec{J}_V = \frac{e\vec{B}}{2\pi^2}\mu_A, {\rm for}\, \mu_A\neq 0,
\end{equation}
where $\mu_A$ is the axial chemical potential that quantifies the axial charge 
asymmetry or imbalance between right- and left-handed quarks in the 
plasma \cite{Fukushima:2008xe,Son:2009tf,Zakharov:2012vv,Fukushima:2012vr}. 

The interplay between the CSE and CME in the QGP produced in heavy ion collisions, can lead to the 
production of a gapless collective mode -- termed the chiral magnetic wave (CMW) \cite{Kharzeev:2010gd}, 
stemming from the coupling between the density waves of the electric and chiral charges.  
The propagation of the CMW is sustained by alternating oscillations of the local electric and 
chiral charge densities that feed into each other to ultimately transport positive (negative)
charges out-of-plane and negative (positive) charges in-plane to form an electric quadrupole. 
Here, the reaction plane $\Psi_{\rm RP}$, is defined by the impact vector $\vec{b}$ and the beam direction,
so the poles of the quadrupole lie along the direction of the $\vec{B}$-field (out-of-plane) which is 
essentially perpendicular to $\Psi_{\rm RP}$. 

The electric charge quadrupole can induce charge-dependent quadrupole 
correlations between the positively- and negatively-charged particles produced in 
the collisions \cite{Kharzeev:2010gd,Liao:2014ava,Huang:2015oca,Kharzeev:2015znc,
Stephanov:2013tga,Han:2019fce,Zhao:2019ybo}. 
Such correlations can be measured with suitable correlators to aid full characterization of the CMW. 

A pervasive approach employed in prior, as well as ongoing experimental studies of the CMW, is  to 
measure the elliptic- or quadrupole flow difference between negatively- and positively 
charged particles \cite{Burnier:2011bf,Burnier:2012ae}:
\begin{eqnarray} \label{eq:2}
\Delta v_{2}&  \equiv & v_{2}^{-} - v_{2}^{+} \simeq rA_{\rm ch}, \nonumber \\
 A_{\rm ch}& = &\frac{(N^{+} -N^{-})}{(N^{+} +N^{-})} 
\end{eqnarray}
as a function of charge asymmetry $A_{\rm ch}$. Here, 
$N^{\pm}$ denotes the number of positively- (negatively-) charged hadrons measured in 
a given event; the slope parameter $r$, which is experimentally determined from the measurements,
is purported to give an estimate of the strength of the CMW signal
 \cite{Kharzeev:2010gd,Liao:2014ava,Voloshin:2014gja,Adam:2015vje,Huang:2015oca,Kharzeev:2015znc,Zhao:2019ybo}.
%
%
\begin{figure*}[t]
\begin{center}
\includegraphics[width=0.80\linewidth, angle=0]{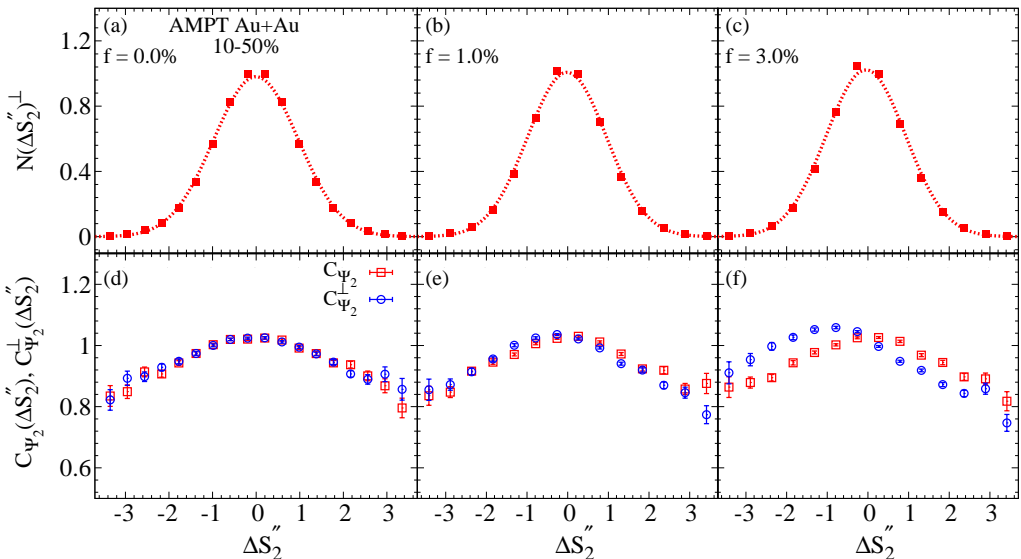}
\vskip -0.25in
\caption{ Simulated $N(\Delta S^{''}_{2})^{\bot}$ distributions (with respect to $\Psi_{\rm 2}$) 
for several input values of quadrupole charge separation characterized by $f_q$  (a-c); comparison of the  
$C_{\Psi_2}(\Delta S^{''}_{2})$ and $C_{\Psi_2}^{\perp}(\Delta S^{''}_{2})$ correlation functions 
for the same values of $f_q$ (d-f). The simulated results are for 10-50\% Au+Au 
collisions at $\sqrtsNN~=~200$ GeV.  
}
\label{fig1} 
\end{center}
\vskip -0.3in
\end{figure*} 
%
%
However, a wealth of measurements reported by the ALICE \cite{Voloshin:2014gja,Adam:2015vje}, 
CMS \cite{SangEonParkonbehalfoftheCMS:2017ams,Sirunyan:2017tax} and 
STAR \cite{Adamczyk:2015eqo,Shou:2018zvw} collaborations, highlight a significant influence from 
the effects of background, suggesting a need for supplemental measurements with improved 
correlators that not only suppress background, but are also  sensitive to small CMW signals in the 
presence of these backgrounds.



In prior work, we have proposed \cite{Magdy:2017yje} and validated the 
utility \cite{Magdy:2018lwk,Huang:2019vfy,Magdy:2020wiu} of the  
$R_{\Psi_m}(\Delta S)$ correlator for robust detection and characterization of the CME-driven
dipole charge separation relative to the $\Psi_{2,3}$ planes. Here, we follow the lead of 
Ref.~\cite{Shen:2019puh} by first, extending the correlator for study of the  
CMW-driven quadrupole charge separation, followed by detailed sensitivity tests of the correlator with 
the aid of  AMPT model simulations. 


The extended correlators, $R^{(d)}_{\Psi_m}(\Delta S_{d})$, are constructed for each 
event plane $\Psi_m$, as the ratio:
\be
R^{(d)}_{\Psi_m}(\Delta S_{d}) = C_{\Psi_m}(\Delta S_{d})/C_{\Psi_m}^{\perp}(\Delta S_{d}), \, m=2,3,
\label{eq:4}
\ee
where $d=$ 1 and 2 denote dipole and quadrupole charge separation respectively, and 
$C_{\Psi_m}(\Delta S_{d})$ and $C_{\Psi_m}^{\perp}(\Delta S_{d})$ are correlation functions 
designed to quantify the dipole and quadrupole charge separation $\Delta S_{d}$, parallel 
and perpendicular (respectively) to the $\vec{B}$-field, i.e., perpendicular 
and parallel (respectively) to $\mathrm{\Psi_{RP}}$.

The correlation functions used to quantify the dipole and quadrupole charge separation parallel to the $\vec{B}$-field, 
are constructed from the ratio of two distributions: 
\be
C_{\Psi_{m}}(\Delta S_{d}) = \frac{N_{\text{real}}(\Delta S_{d})}{N_{\text{Shuffled}}(\Delta S_{d})}, \, m=2,3,
\label{eq:5}
\ee
where $N_{\text{real}}(\Delta S_{d})$ is the distribution over events, of charge separation relative to the $\Psi_m$ planes in each event:
\be
\Delta S_{d} = \frac{{\sum\limits_1^p {\sin (\frac{m^{d}}{2}\Delta {\varphi_{m} })} }}{p} - 
\frac{{\sum\limits_1^n {\sin (\frac{m^{d}}{2}\Delta {\varphi_{m}  })} }}{n},
\label{eq:7}
\ee
where $n$ and $p$ are the numbers of negatively- and positively charged hadrons in an event, $\Delta {\varphi_{m}}= \phi - \Psi_{m}$ and $\phi$ is the azimuthal emission angle of the charged hadrons. The $N_{\text{Shuffled}}(\Delta S_{d})$ distribution
is similarly obtained from the same events, following random reassignment (shuffling) of the charge of each particle in an event. This procedure ensures identical properties for the numerator and the denominator in Eq.~\ref{eq:5}, except for the charge-dependent correlations  which are of interest.

The correlation functions $C_{\Psi_{m}}^{\perp}(\Delta S_{d})$, used to quantify the dipole and quadrupole 
charge separation perpendicular to the $\vec{B}$-field, are constructed with the same procedure outlined 
for $C_{\Psi_{m}}(\Delta S_{d})$, but with $\Psi_{m}$ replaced by $\Psi_{m}+\pi/m^{d}$.  Note that 
this rotation of $\Psi_m$ maps the sine terms in Eq.~\ref{eq:7} into cosine terms.

The correlator $R^{(d)}_{\Psi_2}(\Delta S_{d}) = C_{\Psi_2}(\Delta S_{d})/C_{\Psi_2}^{\perp}(\Delta S_{d})$, 
gives a measure of the magnitude of the charge separation (dipole and quadrupole) parallel to 
the $\vec{B}$-field (perpendicular to $\Psi_2$), relative to that for charge separation perpendicular 
to the $\vec{B}$-field (parallel to $\Psi_2$).
Since the CME- and CMW-driven charge separations are strongly correlated with the $\vec{B}$-field direction,  
the correlators $R^{(d)}_{\Psi_3}(\Delta S_{d}) = C_{\Psi_3}(\Delta S_{d})/C_{\Psi_3}^{\perp}(\Delta S_{d})$ 
are insensitive to them, due to the absence of a strong 
correlation between the $\vec{B}$-field and the orientation of the $\Psi_3$ plane. For small systems 
such as $p$/$d$/$^3$He+Au and $p$+Pb,  a similar insensitivity is to be expected for 
$R^{(d)}_{\Psi_2}(\Delta S_{d})$, due to the weak correlation between the $\vec{B}$-field and the 
orientation of the $\Psi_2$ plane. 
For background-driven charge separation however, similar patterns are to be expected for both 
the $R^{(d)}_{\Psi_2}(\Delta S_{d})$ and $R^{(d)}_{\Psi_3}(\Delta S_{d})$ distributions.  

The response and the sensitivity of the $R^{(1)}_{\Psi_2}(\Delta S_{1})$ correlator to CME-driven charge 
separation is detailed in Refs.~\cite{Magdy:2017yje,Magdy:2020wiu}. For CMW-driven charge separation, 
$R^{(2)}_{\Psi_2}(\Delta S_{2})$ is expected to show an approximately linear dependence 
on $\Delta S_{2}$ for $|\Delta S_{2}| \alt 3$, due to a shift in the distributions 
for $C_{\Psi_2}^{\perp}(\Delta S_{d})$ relative to $C_{\Psi_2}(\Delta S_{d})$, induced by the CMW. 
Thus, the slope of the plot of $R^{(2)}_{\Psi_2}(\Delta S_{2})$ vs. $\Delta S_{2}$, encodes 
the magnitude of the  CMW signal. 
This  slope is also influenced by particle number fluctuations and the resolution of the $\Psi_2$
plane which fluctuates about $\Psi_{\rm RP}$. 
The influence of the particle number fluctuations can be minimized by scaling $\Delta S_{2}$ 
by the width $\mathrm{\sigma_{\Delta_{Sh}}}$ of the distribution 
for $N_{\text{shuffled}}(\Delta S_{2})$ {\em i.e.}, $\Delta S_{2}^{'} = \Delta S_{2}/\mathrm{\sigma_{\Delta_{Sh}}}$. 
Similarly, the effects of the event plane resolution can be accounted for by scaling $\Delta S_{2}^{'}$ 
by the resolution factor $\mathrm{\delta_{Res}}$, {\em i.e.},  
$\Delta S_{2}^{''}= \Delta S_{2}^{'}/\mathrm{\delta_{Res}}$, where $\mathrm{\delta_{Res}}$ 
is the event  plane resolution. The efficacy of these scaling factors have been confirmed via 
detailed simulation studies, as well as with data-driven studies. 

Our sensitivity studies for $R^{(2)}_{\Psi_m}(\Delta S_{2})$, 
relative to the $\Psi_{2}$ and $\Psi_{3}$ event planes,  are performed with 
AMPT events in which varying degrees of proxy CMW-driven quadrupole charge separation 
were introduced \cite{Ma:2011uma,Shen:2019puh}. 
The AMPT model is known to give a good representation of the experimentally measured 
particle yields, spectra, flow, etc.,\cite{Lin:2004en,Ma:2016fve,Ma:2013gga,Ma:2013uqa,Bzdak:2014dia,Nie:2018xog}. 
Therefore, it provides a reasonable estimate of both the magnitude and the properties of 
the background-driven quadrupole  charge separation expected in the data collected at RHIC and the LHC.

%

 We simulated Au+Au collisions at $\sqrtsNN~=~200$ GeV with the same AMPT model version 
used in our prior studies \cite{Huang:2019vfy,Shen:2019puh,Magdy:2020wiu}; this version 
incorporates both string melting and local charge conservation.
In brief, the model follows four primary stages:
(i) an initial-state, (ii) a parton cascade phase, 
(iii) a hadronization phase in which partons are converted to hadrons, and 
(iv) a hadronic re-scattering phase. 
The initial-state essentially simulates the spatial and momentum distributions of mini-jet partons 
from QCD  hard processes and soft string excitations as encoded in the HIJING 
model~\cite{Wang:1991hta,Gyulassy:1994ew}. The parton cascade considers 
the strong interactions among partons via elastic partonic collisions~\cite{Zhang:1997ej}. 
Hadronization is simulated via a coalescence mechanism. After hadronization, the ART model 
is invoked to simulate baryon-baryon, baryon-meson and meson-meson interactions~\cite{Li:1995pra}.  

 A formal mechanism for generation of the CMW is not implemented in the AMPT model. 
However,  a proxy  CMW-induced quadrupole 
charge separation can be implemented~\cite{Ma:2011uma,Ma:2014iva} by interchanging the the position coordinates ($x$, $y$, $z$) for a fraction ($f_q$) of the in-plane light quarks ($u$, $d$ and $s$) carrying positive (negative) charges with  out-of-plane quarks carrying negative (positive) charges, at the start of the partonic stage.
This procedure lends itself to two quadrupole charge configurations, relative to the 
in-plane and out-of-plane orientations. The first or Type (I), is for events with  negative net charge ($A_{\rm ch}<-0.01$) 
in which the $u$ and $\bar{d}$ are set to  be concentrated on the equator of the quadrupole (in-plane),
while $\bar{d}$ and $u$ quarks are set to be concentrated at the poles of the quadrupole (out-of-plane).
The second or Type (II), is for events with  positive net charge  ($A_{\rm ch}>-0.01$) in which the 
in-plane and out-of-plane quark configurations are swapped. The latter configuration was employed for 
the bulk of the AMPT events generated with proxy input signals. The magnitude of the proxy CMW signal is set by the 
fraction $f_q$, which serves to characterize the strength of the quadrupole charge separation.

The AMPT events with  varying degrees of proxy CMW signals were analyzed with the $R^{(2)}_{\Psi_{2,3}}(\Delta S_{2})$
correlators to identify and quantify their response to the respective input signals, following the requisite 
corrections for particle number fluctuations ($\Delta S_{2}^{'} = \Delta S_{2}/\mathrm{\sigma_{\Delta_{Sh}}}$) and 
event-plane resolution ($\Delta S_{2}^{''}= \Delta S_{2}^{'}/\mathrm{\delta_{Res}}$), as described earlier.

%
%
\begin{figure*}[th]
\begin{center}
\includegraphics[width=0.80\linewidth, angle=0]{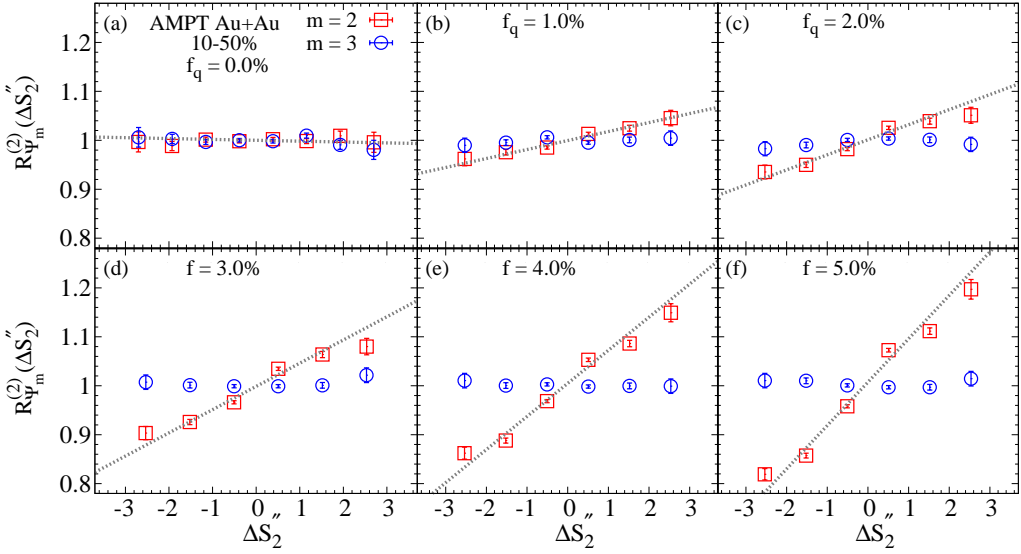}
\vskip -0.25in
\caption{$R^{(2)}_{\Psi_m}(\Delta S_{2})$ vs. $\Delta S^{''}$  for several input values of quadrupole charge separation 
characterized by $f_q$, for 10-50\% Au+Au collisions ($\sqrtsNN~=~200$ GeV).
} 
\label{fig2} 
\end{center}
\vskip -0.3in
\end{figure*} 
%
%
%
\begin{figure}[t]
\begin{center}
\vskip -0.15in
\includegraphics[width=1.05\linewidth, angle=-00]{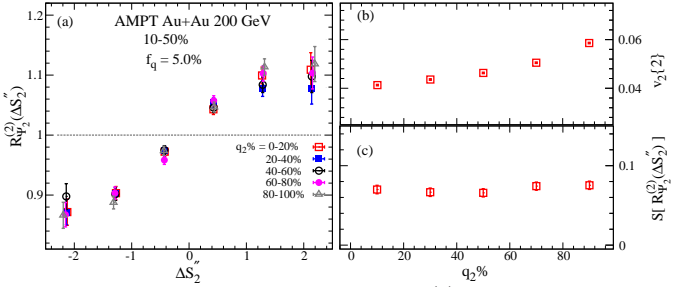}
\vskip -0.15in
\caption{ 
Comparison of the simulated $R^{(2)}_{\Psi_2}(\Delta S^{2})$ correlators for $q_2$ selected events 
in $10-50$\% central, Au+Au collisions at $\sqrtsNN = 200$~GeV (a); $v_2(q_2)$ vs. $q_2$ for  
the same $q_2$-selected events. Panel (c) shows a comparison of the slopes extracted from 
$R^{(2)}_{\Psi_2}$ vs.  $\Delta S^{''}_{2}$ distributions shown in panel (a).
} 
\label{fig4} 
\end{center}
\vskip -0.20in
\end{figure} 
%
%
%
\begin{figure}[t]
\vskip -0.15in
\begin{center}
\includegraphics[width=0.65\linewidth, angle=-90]{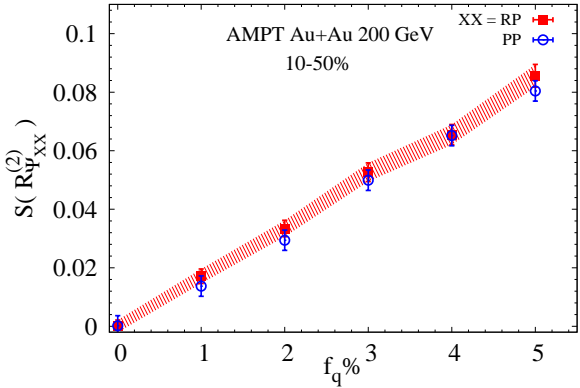}
\vskip -0.15in
\caption{  $f_q$ dependence of the slopes extracted from the $R^{(2)}_{\Psi_2}(\Delta S^{''}_{2})$ vs. $\Delta S^{''}_{2}$
distributions. Results are shown for 10-50\% central Au+Au ($\sqrtsNN~=~200$ GeV) AMPT events.
} \label{fig3} 
\end{center}
\vskip -0.3in
\end{figure} 

The top panels of Fig. \ref{fig1} confirm the expected Gaussian distributions for $N(\Delta S^{''}_{2})^{\bot}$,
as well as the shift in its mean value as $f_q$ increases; the mean value is zero for $f_q =0$ (a) and 
progressively shifts to $\Delta S^{''}_{2} <0$ for $f_q > 0$ (b and c).
These CMW-induced shifts for $f_q > 0$, are made more transparent in Figs. \ref{fig1} (d)-(f) where the 
shift of $C_{\Psi_2}^{\perp}(\Delta S^{''}_{2})$ relative to the $C_{\Psi_2}(\Delta S^{''}_{2})$ correlation function 
is apparent c.f. Fig. \ref{fig1} (f).

The $R^{(2)}_{\Psi_2}(\Delta S^{''}_{2})$ and $R^{(2)}_{\Psi_3}(\Delta S^{''}_{2})$ correlators, obtained for several 
input values of $f_q$, are shown in Fig.~\ref{fig2}. They indicate an essentially flat distribution for 
$R^{(2)}_{\Psi_3}(\Delta S^{''}_{2})$ irrespective of the value of $f_q$. These patterns are consistent with the expected 
insensitivity of  $R^{(2)}_{\Psi_3}(\Delta S^{''}_{2})$ to CMW-driven charge separation due to the absence of a strong 
correlation between the $\vec{B}$-field and the orientation of the $\Psi_3$ plane.
Figs.~\ref{fig2} (a)-(f) show that the $R^{(2)}_{\Psi_2}(\Delta S^{''}_{2})$ correlator evolves from a 
flat distribution for $f_q=0$, to an approximately linear dependence on $\Delta S^{''}_{2}$ 
(for $|\Delta S^{''}_{2}| \alt 3$) with slopes that reflect the increase in the magnitude of 
the input CMW-driven charge separation with $f_q$. These patterns not only confirm the 
input quadrupole charge separation signal in each case; they suggest that the $R^{(2)}_{\Psi_m}(\Delta S_{2})$ 
correlator is relatively insensitive to a possible $v_{2,3}$-driven background [and their associated fluctuations] 
as well as the local charge conservation effects implemented in the AMPT model.
Note the essentially flat distributions for $R^{(2)}_{\Psi_3}(\Delta S^{''}_{2})$ and 
for $R^{(2)}_{\Psi_2}(\Delta S^{''}_{2})$ when the input signal is set to zero.

This insensitivity can be further checked via the event-shape engineering, through fractional cuts 
on the distribution of the magnitude of the $q_2$ flow vector~\cite{Schukraft:2012ah}. 
Here, the underlying notion is that elliptic flow ${v_2}$, which is a major driver 
of background correlations, is strongly correlated with $q_2$~\cite{Acharya:2017fau,Zhao:2018ixy}. 
Thus, the magnitude of the background correlations can be increased(decreased) by selecting events 
with larger(smaller) $q_2$ values. Such selections were made by splitting each event 
into three sub-events; $A[\eta < -0.3]$, $B[|\eta| < 0.4]$, and $C[\eta > 0.3]$, where sub-event 
$B$ was used to evaluate $q_2$, and the other sub-events used to evaluate 
$R^{(2)}_{\Psi_2}(\Delta S^{''}_{2})$ via the methods described earlier.

Figure~\ref{fig4} shows a comparison of the $q_2$-selected  
$R^{(2)}_{\Psi_{2}}$ distributions (a),  $v_2$ (b) and the slopes (c) extracted 
from the distributions shown in panel (a), respectively.  These results were obtained 
for 10-50\% central Au+Au collisions with $f_q$=5\%. They indicate that while $v_2$ increases  
with $q_2$, the corresponding slope for the $R^{(2)}_{\Psi_{2}}$ 
correlators (Fig.~\ref{fig4} (c)) show little, if any, change.  
This insensitivity to the value of $q_2$ is incompatible with a dominating influence 
of background-driven contributions to $R^{(2)}_{\Psi_2}(\Delta S^{''}_{2})$.
It is noteworthy that a further analysis performed for background-driven charge separation with strong local 
charge conservation, also indicated that $R^{(2)}_{\Psi_2}(\Delta S^{''}_{2})$ is essentially 
insensitive to this background.

The $R^{(2)}_{\Psi_2}(\Delta S^{''}_{2})$ distributions shown in Fig.~\ref{fig2}, 
indicate slopes that visibly increase with $f_q$. To quantify the measured signal strengths, 
we extracted the slope $S$, of the respective $R^{(2)}_{\Psi_2}(\Delta S^{''}_{2})$ distributions 
shown in the figure.  Fig.~\ref{fig3} indicates a linear dependence of these slopes 
on $f_q$. It also shows that the magnitude and trends of $S$ are independent of the event plane 
used in the analysis. These results suggests that the $R^{(2)}_{\Psi_2}$ correlator not only suppresses 
background, but is sensitive to small CMW-driven charge separation in the presence of such backgrounds.
%
%
%
\begin{figure}[t]
\vskip -0.15in
\begin{center}
\includegraphics[width=0.85\linewidth, angle=-0]{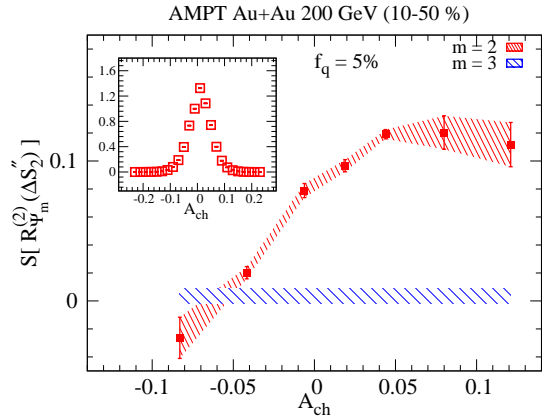}
\vskip -0.15in
\caption{ $A_{\rm ch}$ dependence of the slopes extracted from the $R^{(2)}_{\Psi_2}(\Delta S^{''}_{2})$ vs. $\Delta S^{''}_{2}$
distributions for different $A_{\rm ch}$ selections. The inset shows a normalized distribution 
of  $A_{\rm ch}$. Results are shown for 10-50\% central Au+Au ($\sqrtsNN~=~200$ GeV) AMPT events.
} \label{fig5} 
\end{center}
\vskip -0.3in
\end{figure} 

The slopes of the $R^{(2)}_{\Psi_2}(\Delta S^{''}_{2})$ vs. $\Delta S^{''}_{2}$ distributions can also be 
explored as a function of the charge asymmetry $A_{\rm ch}$ as shown in Fig.~\ref{fig5}. Here, the $A_{\rm ch}$ 
distribution shown  in the inset, hints at the fact that the model parameters used  in the AMPT simulations 
were chosen to give a positive net charge, when averaged over all events. 
Fig.~\ref{fig5} shows the expected decrease of $S$ with $A_{\rm ch}$ for $A_{\rm ch} < 0$. 
It also shows that the sign of $S$ can even be flipped for sufficiently large negative values of $A_{\rm ch}$,
in accord with expectations. Fig.~\ref{fig5}  also shows that the slopes 
for $R^{(2)}_{\Psi_3}(\Delta S^{''}_{2})$ vs. $\Delta S^{''}_{2}$ are insensitive to $A_{\rm ch}$ as might 
be expected. These dependencies could serve as further aids to CMW signal detection and characterization 
in experimental measurements.
 
In summary, we have extended the $R^{(1)}_{\Psi_{m}}(\Delta S_{1})$ correlator, 
previously used to measure CME-induced dipole charge separation, to include the
$R^{(2)}_{\Psi_{m}}(\Delta S_{2})$ correlator, which can be used to measure 
CMW-driven quadrupole charge separation. Validation tests involving varing degrees of 
proxy CMW signals injected into AMPT events, show that the $R^{(2)}_{\Psi_{m}}(\Delta S_{2})$ correlator 
provides discernible responses for background- and CMW-driven charge separation which could aid
robust identification of the CMW.  They also indicate a level of sensitivity that would allow for a robust 
experimental characterization of the purported CMW signals via $R^{(2)}_{\Psi_{m}}(\Delta S_{2})$
measurements in heavy-ion collisions.

\section*{Acknowledgments}
\begin{acknowledgments}
This research is supported by the US Department of Energy, Office of Science, Office of Nuclear Physics, 
under contracts DE-FG02-87ER40331.A008  (RL), DE-FG02-94ER40865 (NM) and 
by the National Natural Science Foundation of China under Grants No. 11890714, No. 11835002, No. 11961131011, No. 11421505, the Key Research Program of the Chinese Academy of Sciences under Grant No. XDPB09 (L.H. and G.-L.M.)


%
\end{acknowledgments}
%
%
\bibliography{lpvpub} 

\begin{thebibliography}{57}%
\makeatletter
\providecommand \@ifxundefined [1]{%
 \@ifx{#1\undefined}
}%
\providecommand \@ifnum [1]{%
 \ifnum #1\expandafter \@firstoftwo
 \else \expandafter \@secondoftwo
 \fi
}%
\providecommand \@ifx [1]{%
 \ifx #1\expandafter \@firstoftwo
 \else \expandafter \@secondoftwo
 \fi
}%
\providecommand \natexlab [1]{#1}%
\providecommand \enquote  [1]{``#1''}%
\providecommand \bibnamefont  [1]{#1}%
\providecommand \bibfnamefont [1]{#1}%
\providecommand \citenamefont [1]{#1}%
\providecommand \href@noop [0]{\@secondoftwo}%
\providecommand \href [0]{\begingroup \@sanitize@url \@href}%
\providecommand \@href[1]{\@@startlink{#1}\@@href}%
\providecommand \@@href[1]{\endgroup#1\@@endlink}%
\providecommand \@sanitize@url [0]{\catcode `\\12\catcode `\$12\catcode
  `\&12\catcode `\#12\catcode `\^12\catcode `\_12\catcode `\%12\relax}%
\providecommand \@@startlink[1]{}%
\providecommand \@@endlink[0]{}%
\providecommand \url  [0]{\begingroup\@sanitize@url \@url }%
\providecommand \@url [1]{\endgroup\@href {#1}{\urlprefix }}%
\providecommand \urlprefix  [0]{URL }%
\providecommand \Eprint [0]{\href }%
\providecommand \doibase [0]{http://dx.doi.org/}%
\providecommand \selectlanguage [0]{\@gobble}%
\providecommand \bibinfo  [0]{\@secondoftwo}%
\providecommand \bibfield  [0]{\@secondoftwo}%
\providecommand \translation [1]{[#1]}%
\providecommand \BibitemOpen [0]{}%
\providecommand \bibitemStop [0]{}%
\providecommand \bibitemNoStop [0]{.\EOS\space}%
\providecommand \EOS [0]{\spacefactor3000\relax}%
\providecommand \BibitemShut  [1]{\csname bibitem#1\endcsname}%
\let\auto@bib@innerbib\@empty
\bibitem [{\citenamefont {Kharzeev}(2006)}]{Kharzeev:2004ey}%
  \BibitemOpen
  \bibfield  {author} {\bibinfo {author} {\bibfnamefont {Dmitri}\ \bibnamefont
  {Kharzeev}},\ }\bibfield  {title} {\enquote {\bibinfo {title} {{Parity
  violation in hot QCD: Why it can happen, and how to look for it}},}\ }\href
  {\doibase 10.1016/j.physletb.2005.11.075} {\bibfield  {journal} {\bibinfo
  {journal} {Phys. Lett.}\ }\textbf {\bibinfo {volume} {B633}},\ \bibinfo
  {pages} {260--264} (\bibinfo {year} {2006})},\ \Eprint
  {http://arxiv.org/abs/hep-ph/0406125} {arXiv:hep-ph/0406125} \BibitemShut
  {NoStop}%
\bibitem [{\citenamefont {Liao}(2015)}]{Liao:2014ava}%
  \BibitemOpen
  \bibfield  {author} {\bibinfo {author} {\bibfnamefont {Jinfeng}\ \bibnamefont
  {Liao}},\ }\bibfield  {title} {\enquote {\bibinfo {title} {{Anomalous
  transport effects and possible environmental symmetry ‘violation’ in
  heavy-ion collisions}},}\ }\href {\doibase 10.1007/s12043-015-0984-x}
  {\bibfield  {journal} {\bibinfo  {journal} {Pramana}\ }\textbf {\bibinfo
  {volume} {84}},\ \bibinfo {pages} {901--926} (\bibinfo {year} {2015})},\
  \Eprint {http://arxiv.org/abs/1401.2500} {arXiv:1401.2500 [hep-ph]}
  \BibitemShut {NoStop}%
\bibitem [{\citenamefont {Miransky}\ and\ \citenamefont
  {Shovkovy}(2015)}]{Miransky:2015ava}%
  \BibitemOpen
  \bibfield  {author} {\bibinfo {author} {\bibfnamefont {Vladimir~A.}\
  \bibnamefont {Miransky}}\ and\ \bibinfo {author} {\bibfnamefont {Igor~A.}\
  \bibnamefont {Shovkovy}},\ }\bibfield  {title} {\enquote {\bibinfo {title}
  {{Quantum field theory in a magnetic field: From quantum chromodynamics to
  graphene and Dirac semimetals}},}\ }\href {\doibase
  10.1016/j.physrep.2015.02.003} {\bibfield  {journal} {\bibinfo  {journal}
  {Phys. Rept.}\ }\textbf {\bibinfo {volume} {576}},\ \bibinfo {pages} {1--209}
  (\bibinfo {year} {2015})},\ \Eprint {http://arxiv.org/abs/1503.00732}
  {arXiv:1503.00732 [hep-ph]} \BibitemShut {NoStop}%
\bibitem [{\citenamefont {Huang}(2016)}]{Huang:2015oca}%
  \BibitemOpen
  \bibfield  {author} {\bibinfo {author} {\bibfnamefont {Xu-Guang}\
  \bibnamefont {Huang}},\ }\bibfield  {title} {\enquote {\bibinfo {title}
  {{Electromagnetic fields and anomalous transports in heavy-ion collisions ---
  A pedagogical review}},}\ }\href {\doibase 10.1088/0034-4885/79/7/076302}
  {\bibfield  {journal} {\bibinfo  {journal} {Rept. Prog. Phys.}\ }\textbf
  {\bibinfo {volume} {79}},\ \bibinfo {pages} {076302} (\bibinfo {year}
  {2016})},\ \Eprint {http://arxiv.org/abs/1509.04073} {arXiv:1509.04073
  [nucl-th]} \BibitemShut {NoStop}%
\bibitem [{\citenamefont {Kharzeev}\ \emph {et~al.}(2016)\citenamefont
  {Kharzeev}, \citenamefont {Liao}, \citenamefont {Voloshin},\ and\
  \citenamefont {Wang}}]{Kharzeev:2015znc}%
  \BibitemOpen
  \bibfield  {author} {\bibinfo {author} {\bibfnamefont {D.~E.}\ \bibnamefont
  {Kharzeev}}, \bibinfo {author} {\bibfnamefont {J.}~\bibnamefont {Liao}},
  \bibinfo {author} {\bibfnamefont {S.~A.}\ \bibnamefont {Voloshin}}, \ and\
  \bibinfo {author} {\bibfnamefont {G.}~\bibnamefont {Wang}},\ }\bibfield
  {title} {\enquote {\bibinfo {title} {{Chiral magnetic and vortical effects in
  high-energy nuclear collisions—A status report}},}\ }\href {\doibase
  10.1016/j.ppnp.2016.01.001} {\bibfield  {journal} {\bibinfo  {journal} {Prog.
  Part. Nucl. Phys.}\ }\textbf {\bibinfo {volume} {88}},\ \bibinfo {pages}
  {1--28} (\bibinfo {year} {2016})},\ \Eprint {http://arxiv.org/abs/1511.04050}
  {arXiv:1511.04050 [hep-ph]} \BibitemShut {NoStop}%
\bibitem [{\citenamefont {Rogachevskii}\ \emph {et~al.}(2017)\citenamefont
  {Rogachevskii}, \citenamefont {Ruchayskiy}, \citenamefont {Boyarsky},
  \citenamefont {Fröhlich}, \citenamefont {Kleeorin}, \citenamefont
  {Brandenburg},\ and\ \citenamefont {Schober}}]{Rogachevskii:2017uyc}%
  \BibitemOpen
  \bibfield  {author} {\bibinfo {author} {\bibfnamefont {Igor}\ \bibnamefont
  {Rogachevskii}}, \bibinfo {author} {\bibfnamefont {Oleg}\ \bibnamefont
  {Ruchayskiy}}, \bibinfo {author} {\bibfnamefont {Alexey}\ \bibnamefont
  {Boyarsky}}, \bibinfo {author} {\bibfnamefont {Jürg}\ \bibnamefont
  {Fröhlich}}, \bibinfo {author} {\bibfnamefont {Nathan}\ \bibnamefont
  {Kleeorin}}, \bibinfo {author} {\bibfnamefont {Axel}\ \bibnamefont
  {Brandenburg}}, \ and\ \bibinfo {author} {\bibfnamefont {Jennifer}\
  \bibnamefont {Schober}},\ }\bibfield  {title} {\enquote {\bibinfo {title}
  {{Laminar and turbulent dynamos in chiral magnetohydrodynamics-I: Theory}},}\
  }\href {\doibase 10.3847/1538-4357/aa886b} {\bibfield  {journal} {\bibinfo
  {journal} {Astrophys. J.}\ }\textbf {\bibinfo {volume} {846}},\ \bibinfo
  {pages} {153} (\bibinfo {year} {2017})},\ \Eprint
  {http://arxiv.org/abs/1705.00378} {arXiv:1705.00378 [physics.plasm-ph]}
  \BibitemShut {NoStop}%
\bibitem [{\citenamefont {Rubakov}\ and\ \citenamefont
  {Gorbunov}(2017)}]{Gorbunov:2011zz}%
  \BibitemOpen
  \bibfield  {author} {\bibinfo {author} {\bibfnamefont {Valery~A.}\
  \bibnamefont {Rubakov}}\ and\ \bibinfo {author} {\bibfnamefont {Dmitry~S.}\
  \bibnamefont {Gorbunov}},\ }\href {\doibase 10.1142/10447} {\emph {\bibinfo
  {title} {{Introduction to the Theory of the Early Universe}}}}\ (\bibinfo
  {publisher} {World Scientific},\ \bibinfo {address} {Singapore},\ \bibinfo
  {year} {2017})\BibitemShut {NoStop}%
\bibitem [{\citenamefont {Weber}(2005)}]{Weber:2004kj}%
  \BibitemOpen
  \bibfield  {author} {\bibinfo {author} {\bibfnamefont {Fridolin}\
  \bibnamefont {Weber}},\ }\bibfield  {title} {\enquote {\bibinfo {title}
  {{Strange quark matter and compact stars}},}\ }\href {\doibase
  10.1016/j.ppnp.2004.07.001} {\bibfield  {journal} {\bibinfo  {journal} {Prog.
  Part. Nucl. Phys.}\ }\textbf {\bibinfo {volume} {54}},\ \bibinfo {pages}
  {193--288} (\bibinfo {year} {2005})},\ \Eprint
  {http://arxiv.org/abs/astro-ph/0407155} {arXiv:astro-ph/0407155 [astro-ph]}
  \BibitemShut {NoStop}%
\bibitem [{\citenamefont {Vafek}\ and\ \citenamefont
  {Vishwanath}(2014)}]{Vafek:2013mpa}%
  \BibitemOpen
  \bibfield  {author} {\bibinfo {author} {\bibfnamefont {Oskar}\ \bibnamefont
  {Vafek}}\ and\ \bibinfo {author} {\bibfnamefont {Ashvin}\ \bibnamefont
  {Vishwanath}},\ }\bibfield  {title} {\enquote {\bibinfo {title} {{Dirac
  Fermions in Solids: From High-Tc cuprates and Graphene to Topological
  Insulators and Weyl Semimetals}},}\ }\href {\doibase
  10.1146/annurev-conmatphys-031113-133841} {\bibfield  {journal} {\bibinfo
  {journal} {Ann. Rev. Condensed Matter Phys.}\ }\textbf {\bibinfo {volume}
  {5}},\ \bibinfo {pages} {83--112} (\bibinfo {year} {2014})},\ \Eprint
  {http://arxiv.org/abs/1306.2272} {arXiv:1306.2272 [cond-mat.mes-hall]}
  \BibitemShut {NoStop}%
\bibitem [{\citenamefont {Burkov}(2015)}]{Burkov:2015hba}%
  \BibitemOpen
  \bibfield  {author} {\bibinfo {author} {\bibfnamefont {A.~A.}\ \bibnamefont
  {Burkov}},\ }\bibfield  {title} {\enquote {\bibinfo {title} {{Chiral anomaly
  and transport in Weyl metals}},}\ }\href {\doibase
  10.1088/0953-8984/27/11/113201} {\bibfield  {journal} {\bibinfo  {journal}
  {J. Phys. Condens. Matter}\ }\textbf {\bibinfo {volume} {27}},\ \bibinfo
  {pages} {113201} (\bibinfo {year} {2015})},\ \Eprint
  {http://arxiv.org/abs/1502.07609} {arXiv:1502.07609 [cond-mat.mes-hall]}
  \BibitemShut {NoStop}%
\bibitem [{\citenamefont {Gorbar}\ \emph {et~al.}(2018)\citenamefont {Gorbar},
  \citenamefont {Miransky}, \citenamefont {Shovkovy},\ and\ \citenamefont
  {Sukhachov}}]{Gorbar:2017lnp}%
  \BibitemOpen
  \bibfield  {author} {\bibinfo {author} {\bibfnamefont {E.~V.}\ \bibnamefont
  {Gorbar}}, \bibinfo {author} {\bibfnamefont {V.~A.}\ \bibnamefont
  {Miransky}}, \bibinfo {author} {\bibfnamefont {I.~A.}\ \bibnamefont
  {Shovkovy}}, \ and\ \bibinfo {author} {\bibfnamefont {P.~O.}\ \bibnamefont
  {Sukhachov}},\ }\bibfield  {title} {\enquote {\bibinfo {title} {{Anomalous
  transport properties of Dirac and Weyl semimetals (Review Article)}},}\
  }\href {\doibase 10.1063/1.5037551} {\bibfield  {journal} {\bibinfo
  {journal} {Low Temp. Phys.}\ }\textbf {\bibinfo {volume} {44}},\ \bibinfo
  {pages} {487--505} (\bibinfo {year} {2018})},\ \bibinfo {note} {[Fiz. Nizk.
  Temp.44,635(2017)]},\ \Eprint {http://arxiv.org/abs/1712.08947}
  {arXiv:1712.08947 [cond-mat.mes-hall]} \BibitemShut {NoStop}%
\bibitem [{\citenamefont {Adler}(1969)}]{Adler:1969gk}%
  \BibitemOpen
  \bibfield  {author} {\bibinfo {author} {\bibfnamefont {Stephen~L.}\
  \bibnamefont {Adler}},\ }\bibfield  {title} {\enquote {\bibinfo {title}
  {{Axial vector vertex in spinor electrodynamics}},}\ }\href {\doibase
  10.1103/PhysRev.177.2426} {\bibfield  {journal} {\bibinfo  {journal} {Phys.
  Rev.}\ }\textbf {\bibinfo {volume} {177}},\ \bibinfo {pages} {2426--2438}
  (\bibinfo {year} {1969})},\ \bibinfo {note} {[,241(1969)]}\BibitemShut
  {NoStop}%
\bibitem [{\citenamefont {Bell}\ and\ \citenamefont
  {Jackiw}(1969)}]{Bell:1969ts}%
  \BibitemOpen
  \bibfield  {author} {\bibinfo {author} {\bibfnamefont {J.~S.}\ \bibnamefont
  {Bell}}\ and\ \bibinfo {author} {\bibfnamefont {R.}~\bibnamefont {Jackiw}},\
  }\bibfield  {title} {\enquote {\bibinfo {title} {{A PCAC puzzle: $\pi^0 \to
  \gamma \gamma$ in the $\sigma$ model}},}\ }\href {\doibase
  10.1007/BF02823296} {\bibfield  {journal} {\bibinfo  {journal} {Nuovo Cim.}\
  }\textbf {\bibinfo {volume} {A60}},\ \bibinfo {pages} {47--61} (\bibinfo
  {year} {1969})}\BibitemShut {NoStop}%
\bibitem [{\citenamefont {Moore}\ and\ \citenamefont
  {Tassler}(2011)}]{Moore:2010jd}%
  \BibitemOpen
  \bibfield  {author} {\bibinfo {author} {\bibfnamefont {Guy~D.}\ \bibnamefont
  {Moore}}\ and\ \bibinfo {author} {\bibfnamefont {Marcus}\ \bibnamefont
  {Tassler}},\ }\bibfield  {title} {\enquote {\bibinfo {title} {{The Sphaleron
  Rate in SU(N) Gauge Theory}},}\ }\href {\doibase 10.1007/JHEP02(2011)105}
  {\bibfield  {journal} {\bibinfo  {journal} {JHEP}\ }\textbf {\bibinfo
  {volume} {02}},\ \bibinfo {pages} {105} (\bibinfo {year} {2011})},\ \Eprint
  {http://arxiv.org/abs/1011.1167} {arXiv:1011.1167 [hep-ph]} \BibitemShut
  {NoStop}%
\bibitem [{\citenamefont {Mace}\ \emph {et~al.}(2016)\citenamefont {Mace},
  \citenamefont {Schlichting},\ and\ \citenamefont
  {Venugopalan}}]{Mace:2016svc}%
  \BibitemOpen
  \bibfield  {author} {\bibinfo {author} {\bibfnamefont {M.}~\bibnamefont
  {Mace}}, \bibinfo {author} {\bibfnamefont {S.}~\bibnamefont {Schlichting}}, \
  and\ \bibinfo {author} {\bibfnamefont {R.}~\bibnamefont {Venugopalan}},\
  }\bibfield  {title} {\enquote {\bibinfo {title} {{Off-equilibrium sphaleron
  transitions in the Glasma}},}\ }\href {\doibase 10.1103/PhysRevD.93.074036}
  {\bibfield  {journal} {\bibinfo  {journal} {Phys. Rev.}\ }\textbf {\bibinfo
  {volume} {D93}},\ \bibinfo {pages} {074036} (\bibinfo {year} {2016})},\
  \Eprint {http://arxiv.org/abs/1601.07342} {arXiv:1601.07342 [hep-ph]}
  \BibitemShut {NoStop}%
\bibitem [{\citenamefont {Liao}\ \emph {et~al.}(2010)\citenamefont {Liao},
  \citenamefont {Koch},\ and\ \citenamefont {Bzdak}}]{Liao:2010nv}%
  \BibitemOpen
  \bibfield  {author} {\bibinfo {author} {\bibfnamefont {Jinfeng}\ \bibnamefont
  {Liao}}, \bibinfo {author} {\bibfnamefont {Volker}\ \bibnamefont {Koch}}, \
  and\ \bibinfo {author} {\bibfnamefont {Adam}\ \bibnamefont {Bzdak}},\
  }\bibfield  {title} {\enquote {\bibinfo {title} {{On the Charge Separation
  Effect in Relativistic Heavy Ion Collisions}},}\ }\href {\doibase
  10.1103/PhysRevC.82.054902} {\bibfield  {journal} {\bibinfo  {journal} {Phys.
  Rev.}\ }\textbf {\bibinfo {volume} {C82}},\ \bibinfo {pages} {054902}
  (\bibinfo {year} {2010})},\ \Eprint {http://arxiv.org/abs/1005.5380}
  {arXiv:1005.5380 [nucl-th]} \BibitemShut {NoStop}%
\bibitem [{\citenamefont {Koch}\ \emph {et~al.}(2017)\citenamefont {Koch},
  \citenamefont {Schlichting}, \citenamefont {Skokov}, \citenamefont
  {Sorensen}, \citenamefont {Thomas}, \citenamefont {Voloshin}, \citenamefont
  {Wang},\ and\ \citenamefont {Yee}}]{Skokov:2016yrj}%
  \BibitemOpen
  \bibfield  {author} {\bibinfo {author} {\bibfnamefont {Volker}\ \bibnamefont
  {Koch}}, \bibinfo {author} {\bibfnamefont {Soeren}\ \bibnamefont
  {Schlichting}}, \bibinfo {author} {\bibfnamefont {Vladimir}\ \bibnamefont
  {Skokov}}, \bibinfo {author} {\bibfnamefont {Paul}\ \bibnamefont {Sorensen}},
  \bibinfo {author} {\bibfnamefont {Jim}\ \bibnamefont {Thomas}}, \bibinfo
  {author} {\bibfnamefont {Sergei}\ \bibnamefont {Voloshin}}, \bibinfo {author}
  {\bibfnamefont {Gang}\ \bibnamefont {Wang}}, \ and\ \bibinfo {author}
  {\bibfnamefont {Ho-Ung}\ \bibnamefont {Yee}},\ }\bibfield  {title} {\enquote
  {\bibinfo {title} {{Status of the chiral magnetic effect and collisions of
  isobars}},}\ }\href {\doibase 10.1088/1674-1137/41/7/072001} {\bibfield
  {journal} {\bibinfo  {journal} {Chin. Phys.}\ }\textbf {\bibinfo {volume}
  {C41}},\ \bibinfo {pages} {072001} (\bibinfo {year} {2017})},\ \Eprint
  {http://arxiv.org/abs/1608.00982} {arXiv:1608.00982 [nucl-th]} \BibitemShut
  {NoStop}%
\bibitem [{\citenamefont {Joyce}\ and\ \citenamefont
  {Shaposhnikov}(1997)}]{Joyce:1997uy}%
  \BibitemOpen
  \bibfield  {author} {\bibinfo {author} {\bibfnamefont {M.}~\bibnamefont
  {Joyce}}\ and\ \bibinfo {author} {\bibfnamefont {Mikhail~E.}\ \bibnamefont
  {Shaposhnikov}},\ }\bibfield  {title} {\enquote {\bibinfo {title}
  {{Primordial magnetic fields, right-handed electrons, and the Abelian
  anomaly}},}\ }\href {\doibase 10.1103/PhysRevLett.79.1193} {\bibfield
  {journal} {\bibinfo  {journal} {Phys. Rev. Lett.}\ }\textbf {\bibinfo
  {volume} {79}},\ \bibinfo {pages} {1193--1196} (\bibinfo {year} {1997})},\
  \Eprint {http://arxiv.org/abs/astro-ph/9703005} {arXiv:astro-ph/9703005
  [astro-ph]} \BibitemShut {NoStop}%
\bibitem [{\citenamefont {Tashiro}\ \emph {et~al.}(2012)\citenamefont
  {Tashiro}, \citenamefont {Vachaspati},\ and\ \citenamefont
  {Vilenkin}}]{Tashiro:2012mf}%
  \BibitemOpen
  \bibfield  {author} {\bibinfo {author} {\bibfnamefont {Hiroyuki}\
  \bibnamefont {Tashiro}}, \bibinfo {author} {\bibfnamefont {Tanmay}\
  \bibnamefont {Vachaspati}}, \ and\ \bibinfo {author} {\bibfnamefont
  {Alexander}\ \bibnamefont {Vilenkin}},\ }\bibfield  {title} {\enquote
  {\bibinfo {title} {{Chiral Effects and Cosmic Magnetic Fields}},}\ }\href
  {\doibase 10.1103/PhysRevD.86.105033} {\bibfield  {journal} {\bibinfo
  {journal} {Phys. Rev.}\ }\textbf {\bibinfo {volume} {D86}},\ \bibinfo {pages}
  {105033} (\bibinfo {year} {2012})},\ \Eprint {http://arxiv.org/abs/1206.5549}
  {arXiv:1206.5549 [astro-ph.CO]} \BibitemShut {NoStop}%
\bibitem [{\citenamefont {Vilenkin}(1980)}]{Vilenkin:1980fu}%
  \BibitemOpen
  \bibfield  {author} {\bibinfo {author} {\bibfnamefont {A.}~\bibnamefont
  {Vilenkin}},\ }\bibfield  {title} {\enquote {\bibinfo {title} {{EQUILIBRIUM
  PARITY VIOLATING CURRENT IN A MAGNETIC FIELD}},}\ }\href {\doibase
  10.1103/PhysRevD.22.3080} {\bibfield  {journal} {\bibinfo  {journal} {Phys.
  Rev.}\ }\textbf {\bibinfo {volume} {D22}},\ \bibinfo {pages} {3080--3084}
  (\bibinfo {year} {1980})}\BibitemShut {NoStop}%
\bibitem [{\citenamefont {Metlitski}\ and\ \citenamefont
  {Zhitnitsky}(2005)}]{Metlitski:2005pr}%
  \BibitemOpen
  \bibfield  {author} {\bibinfo {author} {\bibfnamefont {Max~A.}\ \bibnamefont
  {Metlitski}}\ and\ \bibinfo {author} {\bibfnamefont {Ariel~R.}\ \bibnamefont
  {Zhitnitsky}},\ }\bibfield  {title} {\enquote {\bibinfo {title} {{Anomalous
  axion interactions and topological currents in dense matter}},}\ }\href
  {\doibase 10.1103/PhysRevD.72.045011} {\bibfield  {journal} {\bibinfo
  {journal} {Phys. Rev.}\ }\textbf {\bibinfo {volume} {D72}},\ \bibinfo {pages}
  {045011} (\bibinfo {year} {2005})},\ \Eprint
  {http://arxiv.org/abs/hep-ph/0505072} {arXiv:hep-ph/0505072 [hep-ph]}
  \BibitemShut {NoStop}%
\bibitem [{\citenamefont {Son}\ and\ \citenamefont
  {Surowka}(2009)}]{Son:2009tf}%
  \BibitemOpen
  \bibfield  {author} {\bibinfo {author} {\bibfnamefont {Dam~T.}\ \bibnamefont
  {Son}}\ and\ \bibinfo {author} {\bibfnamefont {Piotr}\ \bibnamefont
  {Surowka}},\ }\bibfield  {title} {\enquote {\bibinfo {title} {{Hydrodynamics
  with Triangle Anomalies}},}\ }\href {\doibase 10.1103/PhysRevLett.103.191601}
  {\bibfield  {journal} {\bibinfo  {journal} {Phys. Rev. Lett.}\ }\textbf
  {\bibinfo {volume} {103}},\ \bibinfo {pages} {191601} (\bibinfo {year}
  {2009})},\ \Eprint {http://arxiv.org/abs/0906.5044} {arXiv:0906.5044
  [hep-th]} \BibitemShut {NoStop}%
\bibitem [{\citenamefont {Fukushima}\ \emph {et~al.}(2008)\citenamefont
  {Fukushima}, \citenamefont {Kharzeev},\ and\ \citenamefont
  {Warringa}}]{Fukushima:2008xe}%
  \BibitemOpen
  \bibfield  {author} {\bibinfo {author} {\bibfnamefont {Kenji}\ \bibnamefont
  {Fukushima}}, \bibinfo {author} {\bibfnamefont {Dmitri~E.}\ \bibnamefont
  {Kharzeev}}, \ and\ \bibinfo {author} {\bibfnamefont {Harmen~J.}\
  \bibnamefont {Warringa}},\ }\bibfield  {title} {\enquote {\bibinfo {title}
  {{The Chiral Magnetic Effect}},}\ }\href {\doibase
  10.1103/PhysRevD.78.074033} {\bibfield  {journal} {\bibinfo  {journal} {Phys.
  Rev.}\ }\textbf {\bibinfo {volume} {D78}},\ \bibinfo {pages} {074033}
  (\bibinfo {year} {2008})},\ \Eprint {http://arxiv.org/abs/0808.3382}
  {arXiv:0808.3382 [hep-ph]} \BibitemShut {NoStop}%
\bibitem [{\citenamefont {Zakharov}(2012)}]{Zakharov:2012vv}%
  \BibitemOpen
  \bibfield  {author} {\bibinfo {author} {\bibfnamefont {Valentin~I.}\
  \bibnamefont {Zakharov}},\ }\bibfield  {title} {\enquote {\bibinfo {title}
  {{Chiral Magnetic Effect in Hydrodynamic Approximation}},}\ }\href {\doibase
  10.1007/978-3-642-37305-3-11} {\  (\bibinfo {year} {2012}),\
  10.1007/978-3-642-37305-3-11},\ \bibinfo {note} {[Lect. Notes
  Phys.871,295(2013)]},\ \Eprint {http://arxiv.org/abs/1210.2186}
  {arXiv:1210.2186 [hep-ph]} \BibitemShut {NoStop}%
\bibitem [{\citenamefont {Fukushima}(2013)}]{Fukushima:2012vr}%
  \BibitemOpen
  \bibfield  {author} {\bibinfo {author} {\bibfnamefont {Kenji}\ \bibnamefont
  {Fukushima}},\ }\bibfield  {title} {\enquote {\bibinfo {title} {{Views of the
  Chiral Magnetic Effect}},}\ }\href {\doibase 10.1007/978-3-642-37305-3_9}
  {\bibfield  {journal} {\bibinfo  {journal} {Lect. Notes Phys.}\ }\textbf
  {\bibinfo {volume} {871}},\ \bibinfo {pages} {241--259} (\bibinfo {year}
  {2013})},\ \Eprint {http://arxiv.org/abs/1209.5064} {arXiv:1209.5064
  [hep-ph]} \BibitemShut {NoStop}%
\bibitem [{\citenamefont {Kharzeev}\ and\ \citenamefont
  {Yee}(2011)}]{Kharzeev:2010gd}%
  \BibitemOpen
  \bibfield  {author} {\bibinfo {author} {\bibfnamefont {Dmitri~E.}\
  \bibnamefont {Kharzeev}}\ and\ \bibinfo {author} {\bibfnamefont {Ho-Ung}\
  \bibnamefont {Yee}},\ }\bibfield  {title} {\enquote {\bibinfo {title}
  {{Chiral Magnetic Wave}},}\ }\href {\doibase 10.1103/PhysRevD.83.085007}
  {\bibfield  {journal} {\bibinfo  {journal} {Phys. Rev.}\ }\textbf {\bibinfo
  {volume} {D83}},\ \bibinfo {pages} {085007} (\bibinfo {year} {2011})},\
  \Eprint {http://arxiv.org/abs/1012.6026} {arXiv:1012.6026 [hep-th]}
  \BibitemShut {NoStop}%
\bibitem [{\citenamefont {Stephanov}\ and\ \citenamefont
  {Yee}(2013)}]{Stephanov:2013tga}%
  \BibitemOpen
  \bibfield  {author} {\bibinfo {author} {\bibfnamefont {Mikhail}\ \bibnamefont
  {Stephanov}}\ and\ \bibinfo {author} {\bibfnamefont {Ho-Ung}\ \bibnamefont
  {Yee}},\ }\bibfield  {title} {\enquote {\bibinfo {title} {{Charged elliptic
  flow at zero charge asymmetry}},}\ }\href {\doibase
  10.1103/PhysRevC.88.014908} {\bibfield  {journal} {\bibinfo  {journal} {Phys.
  Rev.}\ }\textbf {\bibinfo {volume} {C88}},\ \bibinfo {pages} {014908}
  (\bibinfo {year} {2013})},\ \Eprint {http://arxiv.org/abs/1304.6410}
  {arXiv:1304.6410 [nucl-th]} \BibitemShut {NoStop}%
\bibitem [{\citenamefont {Han}\ and\ \citenamefont {Xu}(2019)}]{Han:2019fce}%
  \BibitemOpen
  \bibfield  {author} {\bibinfo {author} {\bibfnamefont {Zhang-Zhu}\
  \bibnamefont {Han}}\ and\ \bibinfo {author} {\bibfnamefont {Jun}\
  \bibnamefont {Xu}},\ }\bibfield  {title} {\enquote {\bibinfo {title} {{Charge
  asymmetry dependence of the elliptic flow splitting in relativistic heavy-ion
  collisions}},}\ }\href {\doibase 10.1103/PhysRevC.99.044915} {\bibfield
  {journal} {\bibinfo  {journal} {Phys. Rev.}\ }\textbf {\bibinfo {volume}
  {C99}},\ \bibinfo {pages} {044915} (\bibinfo {year} {2019})},\ \Eprint
  {http://arxiv.org/abs/1904.03544} {arXiv:1904.03544 [nucl-th]} \BibitemShut
  {NoStop}%
\bibitem [{\citenamefont {Zhao}\ \emph {et~al.}(2019)\citenamefont {Zhao},
  \citenamefont {Ma},\ and\ \citenamefont {Ma}}]{Zhao:2019ybo}%
  \BibitemOpen
  \bibfield  {author} {\bibinfo {author} {\bibfnamefont {Xin-Li}\ \bibnamefont
  {Zhao}}, \bibinfo {author} {\bibfnamefont {Guo-Liang}\ \bibnamefont {Ma}}, \
  and\ \bibinfo {author} {\bibfnamefont {Yu-Gang}\ \bibnamefont {Ma}},\
  }\bibfield  {title} {\enquote {\bibinfo {title} {{Novel mechanism for
  electric quadrupole moment generation in relativistic heavy-ion
  collisions}},}\ }\href {\doibase 10.1016/j.physletb.2019.04.002} {\bibfield
  {journal} {\bibinfo  {journal} {Phys. Lett.}\ }\textbf {\bibinfo {volume}
  {B792}},\ \bibinfo {pages} {413--418} (\bibinfo {year} {2019})},\ \Eprint
  {http://arxiv.org/abs/1901.04156} {arXiv:1901.04156 [hep-ph]} \BibitemShut
  {NoStop}%
\bibitem [{\citenamefont {Burnier}\ \emph {et~al.}(2011)\citenamefont
  {Burnier}, \citenamefont {Kharzeev}, \citenamefont {Liao},\ and\
  \citenamefont {Yee}}]{Burnier:2011bf}%
  \BibitemOpen
  \bibfield  {author} {\bibinfo {author} {\bibfnamefont {Yannis}\ \bibnamefont
  {Burnier}}, \bibinfo {author} {\bibfnamefont {Dmitri~E.}\ \bibnamefont
  {Kharzeev}}, \bibinfo {author} {\bibfnamefont {Jinfeng}\ \bibnamefont
  {Liao}}, \ and\ \bibinfo {author} {\bibfnamefont {Ho-Ung}\ \bibnamefont
  {Yee}},\ }\bibfield  {title} {\enquote {\bibinfo {title} {{Chiral magnetic
  wave at finite baryon density and the electric quadrupole moment of
  quark-gluon plasma in heavy ion collisions}},}\ }\href {\doibase
  10.1103/PhysRevLett.107.052303} {\bibfield  {journal} {\bibinfo  {journal}
  {Phys. Rev. Lett.}\ }\textbf {\bibinfo {volume} {107}},\ \bibinfo {pages}
  {052303} (\bibinfo {year} {2011})},\ \Eprint {http://arxiv.org/abs/1103.1307}
  {arXiv:1103.1307 [hep-ph]} \BibitemShut {NoStop}%
\bibitem [{\citenamefont {Burnier}\ \emph {et~al.}(2012)\citenamefont
  {Burnier}, \citenamefont {Kharzeev}, \citenamefont {Liao},\ and\
  \citenamefont {Yee}}]{Burnier:2012ae}%
  \BibitemOpen
  \bibfield  {author} {\bibinfo {author} {\bibfnamefont {Y.}~\bibnamefont
  {Burnier}}, \bibinfo {author} {\bibfnamefont {D.~E.}\ \bibnamefont
  {Kharzeev}}, \bibinfo {author} {\bibfnamefont {J.}~\bibnamefont {Liao}}, \
  and\ \bibinfo {author} {\bibfnamefont {H.~U.}\ \bibnamefont {Yee}},\
  }\bibfield  {title} {\enquote {\bibinfo {title} {{From the chiral magnetic
  wave to the charge dependence of elliptic flow}},}\ }\href@noop {} {\
  (\bibinfo {year} {2012})},\ \Eprint {http://arxiv.org/abs/1208.2537}
  {arXiv:1208.2537 [hep-ph]} \BibitemShut {NoStop}%
\bibitem [{\citenamefont {Voloshin}\ and\ \citenamefont
  {Belmont}(2014)}]{Voloshin:2014gja}%
  \BibitemOpen
  \bibfield  {author} {\bibinfo {author} {\bibfnamefont {Sergei~A.}\
  \bibnamefont {Voloshin}}\ and\ \bibinfo {author} {\bibfnamefont {Ronald}\
  \bibnamefont {Belmont}},\ }\bibfield  {title} {\enquote {\bibinfo {title}
  {{Measuring and interpreting charge dependent anisotropic flow}},}\
  }\bibfield  {booktitle} {\emph {\bibinfo {booktitle} {{Proceedings, 24th
  International Conference on Ultra-Relativistic Nucleus-Nucleus Collisions
  (Quark Matter 2014): Darmstadt, Germany, May 19-24, 2014}}},\ }\href
  {\doibase 10.1016/j.nuclphysa.2014.09.030} {\bibfield  {journal} {\bibinfo
  {journal} {Nucl. Phys.}\ }\textbf {\bibinfo {volume} {A931}},\ \bibinfo
  {pages} {992--996} (\bibinfo {year} {2014})},\ \Eprint
  {http://arxiv.org/abs/1408.0714} {arXiv:1408.0714 [nucl-ex]} \BibitemShut
  {NoStop}%
\bibitem [{\citenamefont {Adam}\ \emph {et~al.}(2016)\citenamefont {Adam} \emph
  {et~al.}}]{Adam:2015vje}%
  \BibitemOpen
  \bibfield  {author} {\bibinfo {author} {\bibfnamefont {Jaroslav}\
  \bibnamefont {Adam}} \emph {et~al.} (\bibinfo {collaboration} {ALICE}),\
  }\bibfield  {title} {\enquote {\bibinfo {title} {{Charge-dependent flow and
  the search for the chiral magnetic wave in Pb-Pb collisions at $\sqrt{s_{\rm
  NN}} =$ 2.76 TeV}},}\ }\href {\doibase 10.1103/PhysRevC.93.044903} {\bibfield
   {journal} {\bibinfo  {journal} {Phys. Rev.}\ }\textbf {\bibinfo {volume}
  {C93}},\ \bibinfo {pages} {044903} (\bibinfo {year} {2016})},\ \Eprint
  {http://arxiv.org/abs/1512.05739} {arXiv:1512.05739 [nucl-ex]} \BibitemShut
  {NoStop}%
\bibitem [{\citenamefont {Park}(2017)}]{SangEonParkonbehalfoftheCMS:2017ams}%
  \BibitemOpen
  \bibfield  {author} {\bibinfo {author} {\bibfnamefont {Sang~Eon}\
  \bibnamefont {Park}} (\bibinfo {collaboration} {CMS}),\ }\bibfield  {title}
  {\enquote {\bibinfo {title} {{Charge asymmetry dependence of anisotropic flow
  in pPb and PbPb collisions with the CMS experiment}},}\ }\bibfield
  {booktitle} {\emph {\bibinfo {booktitle} {{Proceedings, 26th International
  Conference on Ultra-relativistic Nucleus-Nucleus Collisions (Quark Matter
  2017): Chicago, Illinois, USA, February 5-11, 2017}}},\ }\href {\doibase
  10.1016/j.nuclphysa.2017.05.084} {\bibfield  {journal} {\bibinfo  {journal}
  {Nucl. Phys.}\ }\textbf {\bibinfo {volume} {A967}},\ \bibinfo {pages}
  {345--348} (\bibinfo {year} {2017})},\ \Eprint
  {http://arxiv.org/abs/1704.06712} {arXiv:1704.06712 [nucl-ex]} \BibitemShut
  {NoStop}%
\bibitem [{\citenamefont {Sirunyan}\ \emph {et~al.}(2019)\citenamefont
  {Sirunyan} \emph {et~al.}}]{Sirunyan:2017tax}%
  \BibitemOpen
  \bibfield  {author} {\bibinfo {author} {\bibfnamefont {Albert~M}\
  \bibnamefont {Sirunyan}} \emph {et~al.} (\bibinfo {collaboration} {CMS}),\
  }\bibfield  {title} {\enquote {\bibinfo {title} {{Probing the chiral magnetic
  wave in pPb and PbPb collisions at $\sqrt{s_{NN}} = 5.02 $TeV using
  charge-dependent azimuthal anisotropies}},}\ }\href {\doibase
  10.1103/PhysRevC.100.064908} {\bibfield  {journal} {\bibinfo  {journal}
  {Phys. Rev.}\ }\textbf {\bibinfo {volume} {C100}},\ \bibinfo {pages} {064908}
  (\bibinfo {year} {2019})},\ \Eprint {http://arxiv.org/abs/1708.08901}
  {arXiv:1708.08901 [nucl-ex]} \BibitemShut {NoStop}%
\bibitem [{\citenamefont {Adamczyk}\ \emph {et~al.}(2015)\citenamefont
  {Adamczyk} \emph {et~al.}}]{Adamczyk:2015eqo}%
  \BibitemOpen
  \bibfield  {author} {\bibinfo {author} {\bibfnamefont {L.}~\bibnamefont
  {Adamczyk}} \emph {et~al.} (\bibinfo {collaboration} {STAR}),\ }\bibfield
  {title} {\enquote {\bibinfo {title} {{Observation of charge asymmetry
  dependence of pion elliptic flow and the possible chiral magnetic wave in
  heavy-ion collisions}},}\ }\href {\doibase 10.1103/PhysRevLett.114.252302}
  {\bibfield  {journal} {\bibinfo  {journal} {Phys. Rev. Lett.}\ }\textbf
  {\bibinfo {volume} {114}},\ \bibinfo {pages} {252302} (\bibinfo {year}
  {2015})},\ \Eprint {http://arxiv.org/abs/1504.02175} {arXiv:1504.02175
  [nucl-ex]} \BibitemShut {NoStop}%
\bibitem [{\citenamefont {Shou}(2019)}]{Shou:2018zvw}%
  \BibitemOpen
  \bibfield  {author} {\bibinfo {author} {\bibfnamefont {Qi-Ye}\ \bibnamefont
  {Shou}} (\bibinfo {collaboration} {STAR}),\ }\bibfield  {title} {\enquote
  {\bibinfo {title} {{Search for the Chiral Magnetic Wave with Anisotropic Flow
  of Identified Particles at RHIC-STAR}},}\ }\bibfield  {booktitle} {\emph
  {\bibinfo {booktitle} {{Proceedings, 27th International Conference on
  Ultrarelativistic Nucleus-Nucleus Collisions (Quark Matter 2018): Venice,
  Italy, May 14-19, 2018}}},\ }\href {\doibase 10.1016/j.nuclphysa.2018.09.016}
  {\bibfield  {journal} {\bibinfo  {journal} {Nucl. Phys.}\ }\textbf {\bibinfo
  {volume} {A982}},\ \bibinfo {pages} {555--558} (\bibinfo {year} {2019})},\
  \Eprint {http://arxiv.org/abs/1809.01980} {arXiv:1809.01980 [nucl-ex]}
  \BibitemShut {NoStop}%
\bibitem [{\citenamefont {Magdy}\ \emph
  {et~al.}(2018{\natexlab{a}})\citenamefont {Magdy}, \citenamefont {Shi},
  \citenamefont {Liao}, \citenamefont {Ajitanand},\ and\ \citenamefont
  {Lacey}}]{Magdy:2017yje}%
  \BibitemOpen
  \bibfield  {author} {\bibinfo {author} {\bibfnamefont {Niseem}\ \bibnamefont
  {Magdy}}, \bibinfo {author} {\bibfnamefont {Shuzhe}\ \bibnamefont {Shi}},
  \bibinfo {author} {\bibfnamefont {Jinfeng}\ \bibnamefont {Liao}}, \bibinfo
  {author} {\bibfnamefont {N.}~\bibnamefont {Ajitanand}}, \ and\ \bibinfo
  {author} {\bibfnamefont {Roy~A.}\ \bibnamefont {Lacey}},\ }\bibfield  {title}
  {\enquote {\bibinfo {title} {{New correlator to detect and characterize the
  chiral magnetic effect}},}\ }\href {\doibase 10.1103/PhysRevC.97.061901}
  {\bibfield  {journal} {\bibinfo  {journal} {Phys. Rev.}\ }\textbf {\bibinfo
  {volume} {C97}},\ \bibinfo {pages} {061901} (\bibinfo {year}
  {2018}{\natexlab{a}})},\ \Eprint {http://arxiv.org/abs/1710.01717}
  {arXiv:1710.01717 [physics.data-an]} \BibitemShut {NoStop}%
\bibitem [{\citenamefont {Magdy}\ \emph
  {et~al.}(2018{\natexlab{b}})\citenamefont {Magdy}, \citenamefont {Shi},
  \citenamefont {Liao}, \citenamefont {Liu},\ and\ \citenamefont
  {Lacey}}]{Magdy:2018lwk}%
  \BibitemOpen
  \bibfield  {author} {\bibinfo {author} {\bibfnamefont {Niseem}\ \bibnamefont
  {Magdy}}, \bibinfo {author} {\bibfnamefont {Shuzhe}\ \bibnamefont {Shi}},
  \bibinfo {author} {\bibfnamefont {Jinfeng}\ \bibnamefont {Liao}}, \bibinfo
  {author} {\bibfnamefont {Peifeng}\ \bibnamefont {Liu}}, \ and\ \bibinfo
  {author} {\bibfnamefont {Roy~A.}\ \bibnamefont {Lacey}},\ }\bibfield  {title}
  {\enquote {\bibinfo {title} {{Examination of the observability of a chiral
  magnetically driven charge-separation difference in collisions of the
  $\mathrm{^{96}_{44}Ru +\, ^{96}_{44}Ru}$ and $\mathrm{^{96}_{40}Zr +\,
  ^{96}_{40}Zr}$ isobars at energies available at the BNL Relativistic Heavy
  Ion Collider}},}\ }\href {\doibase 10.1103/PhysRevC.98.061902} {\bibfield
  {journal} {\bibinfo  {journal} {Phys. Rev.}\ }\textbf {\bibinfo {volume}
  {C98}},\ \bibinfo {pages} {061902} (\bibinfo {year} {2018}{\natexlab{b}})},\
  \Eprint {http://arxiv.org/abs/1803.02416} {arXiv:1803.02416 [nucl-ex]}
  \BibitemShut {NoStop}%
\bibitem [{\citenamefont {Huang}\ \emph {et~al.}(2019)\citenamefont {Huang},
  \citenamefont {Nie},\ and\ \citenamefont {Ma}}]{Huang:2019vfy}%
  \BibitemOpen
  \bibfield  {author} {\bibinfo {author} {\bibfnamefont {Ling}\ \bibnamefont
  {Huang}}, \bibinfo {author} {\bibfnamefont {Mao-Wu}\ \bibnamefont {Nie}}, \
  and\ \bibinfo {author} {\bibfnamefont {Guo-Liang}\ \bibnamefont {Ma}},\
  }\bibfield  {title} {\enquote {\bibinfo {title} {{Sensitivity analysis of the
  chiral magnetic effect observables using a multiphase transport model}},}\
  }\href@noop {} {\  (\bibinfo {year} {2019})},\ \Eprint
  {http://arxiv.org/abs/1906.11631} {arXiv:1906.11631 [nucl-th]} \BibitemShut
  {NoStop}%
\bibitem [{\citenamefont {Magdy}\ \emph {et~al.}(2020)\citenamefont {Magdy},
  \citenamefont {Nie}, \citenamefont {Ma},\ and\ \citenamefont
  {Lacey}}]{Magdy:2020wiu}%
  \BibitemOpen
  \bibfield  {author} {\bibinfo {author} {\bibfnamefont {Niseem}\ \bibnamefont
  {Magdy}}, \bibinfo {author} {\bibfnamefont {Mao-Wu}\ \bibnamefont {Nie}},
  \bibinfo {author} {\bibfnamefont {Guo-Liang}\ \bibnamefont {Ma}}, \ and\
  \bibinfo {author} {\bibfnamefont {Roy~A.}\ \bibnamefont {Lacey}},\ }\bibfield
   {title} {\enquote {\bibinfo {title} {{A sensitivity study of the primary
  correlators used to characterize chiral-magnetically-driven charge
  separation}},}\ }\href@noop {} {\  (\bibinfo {year} {2020})},\ \Eprint
  {http://arxiv.org/abs/2002.07934} {arXiv:2002.07934 [nucl-ex]} \BibitemShut
  {NoStop}%
\bibitem [{\citenamefont {Shen}\ \emph {et~al.}(2019)\citenamefont {Shen},
  \citenamefont {Chen}, \citenamefont {Ma}, \citenamefont {Ma}, \citenamefont
  {Shou}, \citenamefont {Zhang},\ and\ \citenamefont {Zhong}}]{Shen:2019puh}%
  \BibitemOpen
  \bibfield  {author} {\bibinfo {author} {\bibfnamefont {Diyu}\ \bibnamefont
  {Shen}}, \bibinfo {author} {\bibfnamefont {Jinhui}\ \bibnamefont {Chen}},
  \bibinfo {author} {\bibfnamefont {Guoliang}\ \bibnamefont {Ma}}, \bibinfo
  {author} {\bibfnamefont {Yu-Gang}\ \bibnamefont {Ma}}, \bibinfo {author}
  {\bibfnamefont {Qiye}\ \bibnamefont {Shou}}, \bibinfo {author} {\bibfnamefont
  {Song}\ \bibnamefont {Zhang}}, \ and\ \bibinfo {author} {\bibfnamefont
  {Chen}\ \bibnamefont {Zhong}},\ }\bibfield  {title} {\enquote {\bibinfo
  {title} {{Charge asymmetry dependence of flow and a novel correlator to
  detect the chiral magnetic wave in a multiphase transport model}},}\ }\href
  {\doibase 10.1103/PhysRevC.100.064907} {\bibfield  {journal} {\bibinfo
  {journal} {Phys. Rev.}\ }\textbf {\bibinfo {volume} {C100}},\ \bibinfo
  {pages} {064907} (\bibinfo {year} {2019})},\ \Eprint
  {http://arxiv.org/abs/1911.00839} {arXiv:1911.00839 [hep-ph]} \BibitemShut
  {NoStop}%
\bibitem [{\citenamefont {Ma}\ and\ \citenamefont {Zhang}(2011)}]{Ma:2011uma}%
  \BibitemOpen
  \bibfield  {author} {\bibinfo {author} {\bibfnamefont {Guo-Liang}\
  \bibnamefont {Ma}}\ and\ \bibinfo {author} {\bibfnamefont {Bin}\ \bibnamefont
  {Zhang}},\ }\bibfield  {title} {\enquote {\bibinfo {title} {{Effects of final
  state interactions on charge separation in relativistic heavy ion
  collisions}},}\ }\href {\doibase 10.1016/j.physletb.2011.04.057} {\bibfield
  {journal} {\bibinfo  {journal} {Phys. Lett.}\ }\textbf {\bibinfo {volume}
  {B700}},\ \bibinfo {pages} {39--43} (\bibinfo {year} {2011})},\ \Eprint
  {http://arxiv.org/abs/1101.1701} {arXiv:1101.1701 [nucl-th]} \BibitemShut
  {NoStop}%
\bibitem [{\citenamefont {Lin}\ \emph {et~al.}(2005)\citenamefont {Lin},
  \citenamefont {Ko}, \citenamefont {Li}, \citenamefont {Zhang},\ and\
  \citenamefont {Pal}}]{Lin:2004en}%
  \BibitemOpen
  \bibfield  {author} {\bibinfo {author} {\bibfnamefont {Zi-Wei}\ \bibnamefont
  {Lin}}, \bibinfo {author} {\bibfnamefont {Che~Ming}\ \bibnamefont {Ko}},
  \bibinfo {author} {\bibfnamefont {Bao-An}\ \bibnamefont {Li}}, \bibinfo
  {author} {\bibfnamefont {Bin}\ \bibnamefont {Zhang}}, \ and\ \bibinfo
  {author} {\bibfnamefont {Subrata}\ \bibnamefont {Pal}},\ }\bibfield  {title}
  {\enquote {\bibinfo {title} {{A Multi-phase transport model for relativistic
  heavy ion collisions}},}\ }\href {\doibase 10.1103/PhysRevC.72.064901}
  {\bibfield  {journal} {\bibinfo  {journal} {Phys. Rev.}\ }\textbf {\bibinfo
  {volume} {C72}},\ \bibinfo {pages} {064901} (\bibinfo {year} {2005})},\
  \Eprint {http://arxiv.org/abs/nucl-th/0411110} {arXiv:nucl-th/0411110
  [nucl-th]} \BibitemShut {NoStop}%
\bibitem [{\citenamefont {Ma}\ and\ \citenamefont {Lin}(2016)}]{Ma:2016fve}%
  \BibitemOpen
  \bibfield  {author} {\bibinfo {author} {\bibfnamefont {Guo-Liang}\
  \bibnamefont {Ma}}\ and\ \bibinfo {author} {\bibfnamefont {Zi-Wei}\
  \bibnamefont {Lin}},\ }\bibfield  {title} {\enquote {\bibinfo {title}
  {{Predictions for $\sqrt {s_{NN}}=5.02$ TeV Pb+Pb Collisions from a
  Multi-Phase Transport Model}},}\ }\href {\doibase 10.1103/PhysRevC.93.054911}
  {\bibfield  {journal} {\bibinfo  {journal} {Phys. Rev.}\ }\textbf {\bibinfo
  {volume} {C93}},\ \bibinfo {pages} {054911} (\bibinfo {year} {2016})},\
  \Eprint {http://arxiv.org/abs/1601.08160} {arXiv:1601.08160 [nucl-th]}
  \BibitemShut {NoStop}%
\bibitem [{\citenamefont {Ma}(2013)}]{Ma:2013gga}%
  \BibitemOpen
  \bibfield  {author} {\bibinfo {author} {\bibfnamefont {Guo-Liang}\
  \bibnamefont {Ma}},\ }\bibfield  {title} {\enquote {\bibinfo {title}
  {{Decomposition of the jet fragmentation function in high-energy heavy-ion
  collisions}},}\ }\href {\doibase 10.1103/PhysRevC.88.021902} {\bibfield
  {journal} {\bibinfo  {journal} {Phys. Rev.}\ }\textbf {\bibinfo {volume}
  {C88}},\ \bibinfo {pages} {021902} (\bibinfo {year} {2013})},\ \Eprint
  {http://arxiv.org/abs/1306.1306} {arXiv:1306.1306 [nucl-th]} \BibitemShut
  {NoStop}%
\bibitem [{\citenamefont {Ma}(2014{\natexlab{a}})}]{Ma:2013uqa}%
  \BibitemOpen
  \bibfield  {author} {\bibinfo {author} {\bibfnamefont {Guo-Liang}\
  \bibnamefont {Ma}},\ }\bibfield  {title} {\enquote {\bibinfo {title} {{Medium
  modifications of jet shapes in Pb+Pb collisions at $\sqrt{s_{_{\rm NN}}}$ =
  2.76 TeV within a multiphase transport model}},}\ }\href {\doibase
  10.1103/PhysRevC.89.024902} {\bibfield  {journal} {\bibinfo  {journal} {Phys.
  Rev.}\ }\textbf {\bibinfo {volume} {C89}},\ \bibinfo {pages} {024902}
  (\bibinfo {year} {2014}{\natexlab{a}})},\ \Eprint
  {http://arxiv.org/abs/1309.5555} {arXiv:1309.5555 [nucl-th]} \BibitemShut
  {NoStop}%
\bibitem [{\citenamefont {Bzdak}\ and\ \citenamefont
  {Ma}(2014)}]{Bzdak:2014dia}%
  \BibitemOpen
  \bibfield  {author} {\bibinfo {author} {\bibfnamefont {Adam}\ \bibnamefont
  {Bzdak}}\ and\ \bibinfo {author} {\bibfnamefont {Guo-Liang}\ \bibnamefont
  {Ma}},\ }\bibfield  {title} {\enquote {\bibinfo {title} {{Elliptic and
  triangular flow in $p$+Pb and peripheral Pb+Pb collisions from parton
  scatterings}},}\ }\href {\doibase 10.1103/PhysRevLett.113.252301} {\bibfield
  {journal} {\bibinfo  {journal} {Phys. Rev. Lett.}\ }\textbf {\bibinfo
  {volume} {113}},\ \bibinfo {pages} {252301} (\bibinfo {year} {2014})},\
  \Eprint {http://arxiv.org/abs/1406.2804} {arXiv:1406.2804 [hep-ph]}
  \BibitemShut {NoStop}%
\bibitem [{\citenamefont {Nie}\ \emph {et~al.}(2018)\citenamefont {Nie},
  \citenamefont {Huo}, \citenamefont {Jia},\ and\ \citenamefont
  {Ma}}]{Nie:2018xog}%
  \BibitemOpen
  \bibfield  {author} {\bibinfo {author} {\bibfnamefont {Mao-Wu}\ \bibnamefont
  {Nie}}, \bibinfo {author} {\bibfnamefont {Peng}\ \bibnamefont {Huo}},
  \bibinfo {author} {\bibfnamefont {Jiangyong}\ \bibnamefont {Jia}}, \ and\
  \bibinfo {author} {\bibfnamefont {Guo-Liang}\ \bibnamefont {Ma}},\ }\bibfield
   {title} {\enquote {\bibinfo {title} {{Multiparticle azimuthal cumulants in
  $p$+Pb collisions from a multiphase transport model}},}\ }\href {\doibase
  10.1103/PhysRevC.98.034903} {\bibfield  {journal} {\bibinfo  {journal} {Phys.
  Rev.}\ }\textbf {\bibinfo {volume} {C98}},\ \bibinfo {pages} {034903}
  (\bibinfo {year} {2018})},\ \Eprint {http://arxiv.org/abs/1802.00374}
  {arXiv:1802.00374 [hep-ph]} \BibitemShut {NoStop}%
\bibitem [{\citenamefont {Wang}\ and\ \citenamefont
  {Gyulassy}(1991)}]{Wang:1991hta}%
  \BibitemOpen
  \bibfield  {author} {\bibinfo {author} {\bibfnamefont {Xin-Nian}\
  \bibnamefont {Wang}}\ and\ \bibinfo {author} {\bibfnamefont {Miklos}\
  \bibnamefont {Gyulassy}},\ }\bibfield  {title} {\enquote {\bibinfo {title}
  {{HIJING: A Monte Carlo model for multiple jet production in p p, p A and A A
  collisions}},}\ }\href {\doibase 10.1103/PhysRevD.44.3501} {\bibfield
  {journal} {\bibinfo  {journal} {Phys. Rev.}\ }\textbf {\bibinfo {volume}
  {D44}},\ \bibinfo {pages} {3501--3516} (\bibinfo {year} {1991})}\BibitemShut
  {NoStop}%
\bibitem [{\citenamefont {Gyulassy}\ and\ \citenamefont
  {Wang}(1994)}]{Gyulassy:1994ew}%
  \BibitemOpen
  \bibfield  {author} {\bibinfo {author} {\bibfnamefont {Miklos}\ \bibnamefont
  {Gyulassy}}\ and\ \bibinfo {author} {\bibfnamefont {Xin-Nian}\ \bibnamefont
  {Wang}},\ }\bibfield  {title} {\enquote {\bibinfo {title} {{HIJING 1.0: A
  Monte Carlo program for parton and particle production in high-energy
  hadronic and nuclear collisions}},}\ }\href {\doibase
  10.1016/0010-4655(94)90057-4} {\bibfield  {journal} {\bibinfo  {journal}
  {Comput. Phys. Commun.}\ }\textbf {\bibinfo {volume} {83}},\ \bibinfo {pages}
  {307} (\bibinfo {year} {1994})},\ \Eprint
  {http://arxiv.org/abs/nucl-th/9502021} {arXiv:nucl-th/9502021 [nucl-th]}
  \BibitemShut {NoStop}%
\bibitem [{\citenamefont {Zhang}(1998)}]{Zhang:1997ej}%
  \BibitemOpen
  \bibfield  {author} {\bibinfo {author} {\bibfnamefont {Bin}\ \bibnamefont
  {Zhang}},\ }\bibfield  {title} {\enquote {\bibinfo {title} {{ZPC 1.0.1: A
  Parton cascade for ultrarelativistic heavy ion collisions}},}\ }\href
  {\doibase 10.1016/S0010-4655(98)00010-1} {\bibfield  {journal} {\bibinfo
  {journal} {Comput. Phys. Commun.}\ }\textbf {\bibinfo {volume} {109}},\
  \bibinfo {pages} {193--206} (\bibinfo {year} {1998})},\ \Eprint
  {http://arxiv.org/abs/nucl-th/9709009} {arXiv:nucl-th/9709009 [nucl-th]}
  \BibitemShut {NoStop}%
\bibitem [{\citenamefont {Li}\ and\ \citenamefont {Ko}(1995)}]{Li:1995pra}%
  \BibitemOpen
  \bibfield  {author} {\bibinfo {author} {\bibfnamefont {Bao-An}\ \bibnamefont
  {Li}}\ and\ \bibinfo {author} {\bibfnamefont {Che~Ming}\ \bibnamefont {Ko}},\
  }\bibfield  {title} {\enquote {\bibinfo {title} {{Formation of superdense
  hadronic matter in high-energy heavy ion collisions}},}\ }\href {\doibase
  10.1103/PhysRevC.52.2037} {\bibfield  {journal} {\bibinfo  {journal} {Phys.
  Rev.}\ }\textbf {\bibinfo {volume} {C52}},\ \bibinfo {pages} {2037--2063}
  (\bibinfo {year} {1995})},\ \Eprint {http://arxiv.org/abs/nucl-th/9505016}
  {arXiv:nucl-th/9505016 [nucl-th]} \BibitemShut {NoStop}%
\bibitem [{\citenamefont {Ma}(2014{\natexlab{b}})}]{Ma:2014iva}%
  \BibitemOpen
  \bibfield  {author} {\bibinfo {author} {\bibfnamefont {Guo-Liang}\
  \bibnamefont {Ma}},\ }\bibfield  {title} {\enquote {\bibinfo {title} {{Final
  state effects on charge asymmetry of pion elliptic flow in high-energy
  heavy-ion collisions}},}\ }\href {\doibase 10.1016/j.physletb.2014.06.074}
  {\bibfield  {journal} {\bibinfo  {journal} {Phys. Lett.}\ }\textbf {\bibinfo
  {volume} {B735}},\ \bibinfo {pages} {383--386} (\bibinfo {year}
  {2014}{\natexlab{b}})},\ \Eprint {http://arxiv.org/abs/1401.6502}
  {arXiv:1401.6502 [nucl-th]} \BibitemShut {NoStop}%
\bibitem [{\citenamefont {Schukraft}\ \emph {et~al.}(2013)\citenamefont
  {Schukraft}, \citenamefont {Timmins},\ and\ \citenamefont
  {Voloshin}}]{Schukraft:2012ah}%
  \BibitemOpen
  \bibfield  {author} {\bibinfo {author} {\bibfnamefont {Jurgen}\ \bibnamefont
  {Schukraft}}, \bibinfo {author} {\bibfnamefont {Anthony}\ \bibnamefont
  {Timmins}}, \ and\ \bibinfo {author} {\bibfnamefont {Sergei~A.}\ \bibnamefont
  {Voloshin}},\ }\bibfield  {title} {\enquote {\bibinfo {title}
  {{Ultra-relativistic nuclear collisions: event shape engineering}},}\ }\href
  {\doibase 10.1016/j.physletb.2013.01.045} {\bibfield  {journal} {\bibinfo
  {journal} {Phys. Lett.}\ }\textbf {\bibinfo {volume} {B719}},\ \bibinfo
  {pages} {394--398} (\bibinfo {year} {2013})},\ \Eprint
  {http://arxiv.org/abs/1208.4563} {arXiv:1208.4563 [nucl-ex]} \BibitemShut
  {NoStop}%
\bibitem [{\citenamefont {Acharya}\ \emph {et~al.}(2018)\citenamefont {Acharya}
  \emph {et~al.}}]{Acharya:2017fau}%
  \BibitemOpen
  \bibfield  {author} {\bibinfo {author} {\bibfnamefont {Shreyasi}\
  \bibnamefont {Acharya}} \emph {et~al.} (\bibinfo {collaboration} {ALICE}),\
  }\bibfield  {title} {\enquote {\bibinfo {title} {{Constraining the magnitude
  of the Chiral Magnetic Effect with Event Shape Engineering in Pb-Pb
  collisions at $\sqrt{s_\mathrm{NN}}$ = 2.76 TeV}},}\ }\href {\doibase
  10.1016/j.physletb.2017.12.021} {\bibfield  {journal} {\bibinfo  {journal}
  {Phys. Lett.}\ }\textbf {\bibinfo {volume} {B777}},\ \bibinfo {pages}
  {151--162} (\bibinfo {year} {2018})},\ \Eprint
  {http://arxiv.org/abs/1709.04723} {arXiv:1709.04723 [nucl-ex]} \BibitemShut
  {NoStop}%
\bibitem [{\citenamefont {Zhao}(2018)}]{Zhao:2018ixy}%
  \BibitemOpen
  \bibfield  {author} {\bibinfo {author} {\bibfnamefont {Jie}\ \bibnamefont
  {Zhao}},\ }\bibfield  {title} {\enquote {\bibinfo {title} {{Search for the
  Chiral Magnetic Effect in Relativistic Heavy-Ion Collisions}},}\ }\href
  {\doibase 10.1142/S0217751X18300107} {\bibfield  {journal} {\bibinfo
  {journal} {Int. J. Mod. Phys.}\ }\textbf {\bibinfo {volume} {A33}},\ \bibinfo
  {pages} {1830010} (\bibinfo {year} {2018})},\ \Eprint
  {http://arxiv.org/abs/1805.02814} {arXiv:1805.02814 [nucl-ex]} \BibitemShut
  {NoStop}%
\end{thebibliography}%
\end{document}